\documentclass[11pt,a4paper]{article}
\pdfoutput=1

\usepackage{amsmath,amssymb,tikz,hyperref,empheq}
\usepackage{graphicx}
\usepackage{cite}

 \allowdisplaybreaks
\def\be{\begin{equation}}
\def\ee{\end{equation}}
\def\ba#1\ea{\begin{align}#1\end{align}}
\def\bg#1\eg{\begin{gather}#1\end{gather}}
\def\bm#1\em{\begin{multline}#1\end{multline}}
\def\bmd#1\emd{\begin{multlined}#1\end{multlined}}

\def\a{\alpha}
\def\b{\beta}

\def\d{\delta}

\def\e{\epsilon}

\def\G{\Gamma}

\def\m{\mu}
\def\n{\nu}

\def\r{\rho}
\def\s{\sigma}

\def\la{\label}

\def\nn{\nonumber}

\def\({\left(}
\def\){\right)}
\def\[{\left[}
\def\]{\right]}

\def\cA{{\mathcal A}}

\def \be {\begin{equation}}
\def \ee {\end{equation}}
\def \ba {\begin{array}}
\def \ea {\end{array}}
\def \bea{\begin{eqnarray}}
\def \eea{\end{eqnarray}}
\def \nn {\nonumber}

\def \a {\alpha}
\def \b {\beta}

\def \G {\Gamma}
\def \d {\delta}

\def \e {\epsilon}

\def \m {\mu}
\def \n {\nu}

\def \s {\sigma}

\def \r {\rho}

\def \la {\leftarrow}
\def \ra {\rightarrow}

\def\bea{\begin{eqnarray}}
\def\eea{\end{eqnarray}}
\newcommand{\eq}[1]{(\ref{#1})}
\newcommand{\bit}{\begin{itemize}}  \newcommand{\eit}{\end{itemize}}
\newcommand{\ben}{\begin{enumerate}}  \newcommand{\een}{\end{enumerate}}

\def\la{\langle}
\def\ra{\rangle}

\long\def\symbolfootnote[#1]#2{\begingroup%
\def\thefootnote{\fnsymbol{footnote}}\footnote[#1]{#2}\endgroup}

\newcommand{\nthu}{{\it Department of Physics, National Tsing-Hua
  University,
  Hsinchu 30013, Taiwan}}

\newcommand{\ctc}{{\it
Center of Theory and Computation, 
National Tsing-Hua University, Hsinchu 30013, Taiwan}}

\newcommand{\ncts}{{\it  National Center for Theoretical Sciences, Taipei 10617, Taiwan
}}

\newcommand{\sysu}{{\it School of Physics and Astronomy, Sun Yat-Sen University, 2 Daxue Road, Zhuhai 519082, China}}

\begin{document}
\thispagestyle{empty}
\begin{center}

~\vspace{20pt}
  {\Large\bf  Chiral current induced by torsional Weyl anomaly}

\vspace{25pt}

Chong-Sun Chu ${}^{2,3,4}$\symbolfootnote[1]{Email:~\sf
  cschu@phys.nthu.edu.tw}
, Rong-Xin Miao ${}^1$\symbolfootnote[2]{Email:~\sf
  miaorx@mail.sysu.edu.cn}

\vspace{10pt}${{}^{1}}$\sysu \footnote{All the Institutes of authors
  contribute equally to this work, the order of Institutes is adjusted
  for the assessment policy of SYSU.}
  
  \vspace{5pt}${{}^{2}}$\nthu

\vspace{5pt}${{}^{3}}$\ctc

\vspace{5pt}${{}^{4}}$\ncts

\vspace{1cm}

\begin{abstract}
 Torsion can be realized as dislocation in the crystal lattice of
 material. It is particularly interesting if the material has fermions
 in the spectrum, such as graphene, topological insulators, Dirac and
 Weyl semimetals, as it's transport properties can be affected by the
 torsion. In this letter, we find that, due to Weyl anomaly,
 torsion in Dirac and Weyl semimetals can induce novel chiral
 currents, either near a boundary or in a ``conformally flat space''.
 We briefly discuss how to measure this interesting effect in
 experiment.  It is remarkable that these experiments can help to
 clarify the theoretical controversy of whether an imaginary
 Pontryagin density could appear in the Weyl anomaly.
\end{abstract}

\end{center}

\newpage
\setcounter{footnote}{0}
\setcounter{page}{1}

\tableofcontents

\section{Introduction}

The study of anomaly induced transport is an interesting subject (see
\cite{Chernodub:2021nff} for a recent review).
Although anomaly was originally discovered in particle physics,
due to its universal nature, anomaly
has non-trivial implications to a large number of 
physical phenomena ranging over vastly different scales. 
For example, the chiral anomaly of nonabelian gauge theory imposes
nontrivial constraints on the fundamental interaction  of chiral
fermions in the standard model
\cite{Peskin:1995ev}. Chiral anomaly also affects the
transport dynamics of systems with chiral fermions
\cite{Vilenkin:1995um,Vilenkin:1980fu,
 Giovannini:1997eg,alekseev,Fukushima:2012vr,Kharzeev:2007tn,
 Erdmenger:2008rm,Banerjee:2008th, Son:2009tf,  Landsteiner:2011cp}
due to the well-known chiral magnetic and chiral vortical effects
(see \cite{c1,c2,c3} for review).
Interestingly, 
this kind of anomalous transport occurs
only in a material system since nonvanishing chemical potentials
are required. As anomaly itself is intrinsic to the quantum vacuum,
it is an interesting question to ask if anomaly induced
transport can occur
independent of the chemical potentials.

Recently, chiral response associated with chiral anomaly in a torsional
background has been a subject of intensive study.
In general,  a curved spacetime is equipped with a metric which fixes the 
 causal structure
 and metric relations,
 and a connection which defines the parallel
 transport of tensors on the manifold. While the path dependence of
 parallel transport is measured by the curvature, the non-closure of
 parallelism of parallel transport is measure by the torsion.
 For  the parallel transport
 \be
\nabla_\m V^\r = \partial_\m V^\r + \G^\r{}_{\m \n} V^\n,
 \ee
 the torsion
\be \label{torsion}
T^\r{}_{\m\n} := \G^\r{}_{\m\n} - \G^\r{}_{\n\m},
 \ee
is given by the antisymmetric part of the connection.
In Einstein general relativity (GR), the geometry of spacetime
is taken to be torsion free since there is seemingly no
 observational evidence for torsion in the spacetime of our universe
 \cite{Hammond:2002rm}. However, perhaps unexpectedly, torsion finds a
 legitimate position in condensed matter physics since
 torsion appears to be naturally suited for the
 geometrical description of dislocation defects in crystals
 \cite{kondo,kleinert,katanaev,nelson}. Torsion  has been
 realized and studied in diverse material systems such as graphene 
\cite{deJuan:2010zz,Mesaros:2009az}, topological insulators
\cite{Hughes:2011hv,Hughes2013,Hoyos2014}, Dirac and
Weyl semimetal \cite{Chernodub:2015wxa, Sumiyoshi:2015eda,You:2016wbd,
  Huang:2018iys,Ferreiros:2018udw,Nissinen:2019kld}. 
As a result of the specific manner fermion is coupled to torsion,
chiral anomaly could emerge and give rise to
novel chiral response in torsional material systems
\cite{Khaidukov:2018oat,Imaki:2020csc,Amitani:2022xev, Huang:2019haq,
  Nissinen:2019mkw,Laurila:2020yll,Ferreiros:2020uda,Nissinen:2021gke,Valle:2021nfv}.

Just as a system may possess a chiral anomaly which characterizes
the quantum chiral dependence of the system,
generally a system may also possess a Weyl anomaly which characterizes
the quantum scale dependence of the system.
 In general, the Weyl anomaly
is defined as a difference between the trace of renormalized stress tensor
and the renormalized trace of stress tensor \cite{Duff:1993wm,Brown:1976wc}
\be
\cA = \int_M \sqrt{-g}  \Big[
g^{\m\n}\la T_{\m\n} \ra - \la g^{\m\n} T_{\m\n}\ra
  \Big].
\ee
In the presence of a background gauge field, the Weyl anomaly receives
a contribution
\be
\cA = \int_M \sqrt{-g} \; b_1 F_{\m\n} F^{\m\n}
\ee
whose form is universal and is entirely determined by the coefficient $b_1$,
a bulk central charge of the theory.
For the normalization of the gauge field kinetic
term $S= -1/4 \int F^2$, $b_1$ is given by the beta function of the
theory as $b_1 = -\b/2$. 
As Weyl anomaly is independent of the chiral anomaly, it is interesting to
ask if and how it give rises to any transport phenomena in a system. 
The answer is positive. 
Recently, a new kind of induced transport was
discovered  for boundary
vacuum system as a result of the Weyl anomaly.
It was found that
\cite{Chu:2018ksb,Chu:2018ntx}
for any renormalizable  quantum field theory
with a current coupled to an external electromagnetic (EM) field 
\be \label{S}
S_A=\int_M
\sqrt{-g}
  \; J^\m A_\m  ,
  \ee
  the Weyl anomaly
  give rises to an induced magnetization current
  in the vicinity of the boundary
of the vacuum system 
\begin{eqnarray}\label{typeIcurrent}
 \la J_{\mu}\ra = \frac{-2\beta F_{\mu\nu}n^{\nu}}{x} + \cdots, \ x\sim 0.
\end{eqnarray}
Here $x$ is the proper distance to the boundary,
$n_{\mu}$ is the inner normal vector, $...$ denote higher order terms in
$O(x)$ and $\beta$ is the beta function. Hereafter we will drop the
symbol $\la \; \ra$ for the expectation value.
It is instructive  to review the derivation of this result to appreciate how
it could be derived from the Weyl anomaly. In general, for a boundary quantum
field theory,
the renormalized current is generally singular near
the boundary and the expectation value takes the asymptotic form
near $ x \sim 0$:
\begin{eqnarray}\label{current0}
   J_\m = \frac{1}{x^3}  J^{(3)}_{\m}+\frac{1}{x^2}  J^{(2)}_{\m}
  +\frac{1}{x}  J^{(1)}_{\m}
  + J^{(0)} \log x + \cdots,\;\; 
\end{eqnarray}
where $\cdots$ denotes
terms regular at $x=0$, and $J^{(n)}_{\m}$
depend only the background geometry, the background vector field strength
and the type of fields under consideration. 
For current that is conserved
\begin{eqnarray}\label{divergenceless}
  D_\m  J^\m  = 0
\end{eqnarray}
up to  possibly an anomaly term, one can easily  obtain
the gauge invariant solution
\begin{equation}\label{solnJ1}
\begin{split}
  &J^{(3)}_{ \m}=0, \quad \ J^{(2)}_{\m}=0,\\ &J^{(1)}_{\m}=
  \a_1  F_{\m\n} n^\n
  +\a_2 \mathcal{D}_\m k+\a_3 \mathcal{D}_\n k^\n_\m
 +\a_4  \star F_{\m\n}\, n^\n
\end{split}
\end{equation}
where
$F_{\m\n}$, $\star F_{\m\n}$, $n_\m$, $\mathcal{D}_m$, $k_{\m\n}$ and
$h_{\m\n}$ are respectively the background field strength, 
Hodge dual of field strength, 
the normal vector, induced
covariant derivative, extrinsic curvature and induced metric of the
boundary. Now the Weyl anomaly $\cA$ is a function of the background
gauge field. Since it is related to the Logarithmic UV divergent term
of effective action, one can establish the following
``integrability'' relation  \cite{Miao:2017aba,Chu:2018ksb}
\be \label{key}
(\d \cA )_{\partial M_\epsilon} = \Big(\int_{M_\epsilon} dx^4 \sqrt{g} J^\mu \d A_\mu
\Big)_{\log \frac{1}{\epsilon}},
\ee
where  here
a regulator $x \geq \e$ to the boundary is introduced for the integral
on the right hand side (RHS) of \eq{key} and $J^\mu$ is the renormalized
current.
The relation \eq{key}
identifies the boundary contribution of
the variation of the
integrated anomaly $\cal A$
under an arbitrary variation of the gauge field $\d A_\m$ with  the UV
logarithmic divergent part of the integral
involving the expectation value
$J^\m$ of the renormalized $U(1)$ current.
The power of the relation \eq{key} lies in the fact that the left hand
side of \eq{key} is a total variation
and impose
constraints on the RHS of \eq{key} that are
powerful enough to
to fix completely the asymptotic behavior
of the current in terms of the Weyl anomaly of the theory.
Using \eq{key}, one obtain the result \eq{typeIcurrent} for the renormalized
current immediately.
We note that
the  result (\ref{typeIcurrent}) is universal in two remarkable ways.
First, it works for any quantum field theory, and not just conformal field
theory.
Moreover, it is independent of the choices of boundary conditions
since only the bulk central charge,
instead of boundary central charge, appears.
It should be mentioned that the induced magnetization current for
free theories in the vicinity of the 
the boundary
was first obtained by Osborn and 
McAvity in \cite{McAvity:1990we},
the universal result \eq{typeIcurrent}
as well as its intimate relation with the Weyl  anomaly 
were originally established in \cite{Chu:2018ksb}.
Weyl anomaly also has interesting effect in cosmological spacetime.
It was  found that in a conformally flat spacetime $ds^2=
e^{2\sigma}\eta_{\mu\nu}dx^{\mu}dx^{\nu}$ without boundaries, the
 anomalous current is given by \cite{Chernodub:2016lbo,Chernodub:2017jcp}
 \begin{eqnarray}\label{typeIIcurrent}
\la  J^{\mu} \ra =-2\beta F^{ \mu\nu}  \partial_{\nu} \sigma+O(\sigma^2),
\end{eqnarray}
  to the leading order of small $\s$.
  Generalization of the result  \eq{typeIcurrent} to higher dimensions
  and the result  (\ref{typeIIcurrent})
  for arbitrary finite $\s$ can be found in \cite{higher1,higher2} and
  \cite{Zheng:2019xeu}
  respectively. In addition, there is also a novel effect of
  vacuum spin transport induced by electromagnetic field \cite{Chu:2021eae}.

  Central to these results is the fact that
  the anomalous currents \eq{typeIcurrent},
  \eq{typeIIcurrent} emerge as direct response of the Weyl anomaly
  arising from background.
  Motivated by this observation, it is
  natural to expect that similar induced phenomena
  may occur if the fermions are allowed to couple to
  other external backgrounds.
  Now apart from  EM background or spacetime curvature, spacetime
   torsion is  another interesting background to consider.  
   In this paper, we will  study the
   quantum transport phenomena induced by Weyl anomaly in a torsional material.
   In the following, we first
    derive in section 2    the Weyl anomaly for Dirac fermions
coupled to torsion. Using this result,  
we derive the Weyl anomaly induced chiral
and vector currents in section 3.
The discussion is extended to Weyl fermions in section 4. We propose that
measurements of the induced currents
in Weyl semimetal could help to
clarify the theoretical
controversy of whether the Pontryagin density appears in the Weyl anomaly.


\section{Torsion and Weyl Anomaly}
Generally a spacetime is equipped with a metric
and a connection.
 In Einstein general relativity (GR),
 the metricity condition and a symmetric connection
 are adopted so that
 connection is not independent but given by the metric.
In general, departure from GR
is  characterized \cite{Hammond:2002rm}
by the non-metricity tensor
 $Q_{\m\n\r} := \nabla_\m g_{\n\r}$ and the torsion tensor (\ref{torsion})
defined as the antisymmetric part of the connection.
For simplicity, we will focus in this paper the
particular interesting generalization of GR called the Einstein-Cartan
theory where $Q=0$ and the connection is independently
characterized by the torsion tensor.
In this case, the gravitational coupling in 4-dimensions
takes the general form
\begin{eqnarray} \label{action}
  S=\int_M d^4 x \sqrt{-g}\bar{\psi} i \gamma^{\mu} \big(\nabla_{\mu}
  -i  V_{\mu}
  -  i \gamma_5 S_{\mu} \big)
    \psi,
\end{eqnarray}
where $\nabla_{\mu}$ is the covariant derivative defined
with the
standard
Levi-Civita metric connection
and the components of torsion 
\begin{eqnarray} \label{vectors}
  V_{\mu} :=T^{\rho}_{\ \rho\mu}, \quad
  S_{\mu} :=\epsilon_{\mu\nu\rho\sigma} T^{\nu\rho\sigma},
\end{eqnarray}
behave effectively as vectors and axial vectors
\cite{deJuan:2010zz}.
For 3-dimensions, there is no $S_{\mu}$. Instead 
\be
\hat{S} :=\epsilon_{\nu\rho\sigma} T^{\nu\rho\sigma}
\ee
behaves as
pseudo-scalar and the action in 3-dimensions becomes
\cite{deJuan:2010zz,Mesaros:2009az}
\be
S=\int_M d^3x \sqrt{-g}\bar{\psi} i \gamma^{\mu} \big(\nabla_{\mu}
-i  V_{\mu}
- i \gamma_5 \hat{S} \big) \psi.
\ee
We use the
mostly negative convention for the signature of the metric
and the torsion will be taken as a background.

The action \eq{action} is classically Weyl invariant under the local scaling
transformation: 
$\psi \to e^{ -3\s/2} \psi$, $g_{\m\n} \to e^{2\s} g_{\m\n}$,
$V_\m \to V_\m$
and $ S_{\mu} \to S_{\mu}.$
However, quantum mechanically there is an anomaly.
For a manifold with 
a boundary, boundary conditions should be imposed
on half of the spinor components. It can be shown that  
\cite{Vassilevich:2003xt}
Hermicity of the Dirac operator
selects out of the general chiral bag boundary conditions
the following specific ones:
\begin{eqnarray} \label{bagBC}
  (1 \pm i  \gamma^n \gamma^5)\psi|_{\partial M}=0
\end{eqnarray}
where $n$ denote the normal direction.
The one-loop  Weyl anomaly can be obtained by applying the heat kernel
expansion \cite{Vassilevich:2003xt}.
Let us focus on 4-dimensions, the Weyl anomaly reads
\begin{eqnarray} \label{anomaly}
  \mathcal{A}=
  && \frac{1}{24\pi^2}\int_M d^4x\sqrt{-g} \big[F_{\mu\nu}F^{\mu\nu}+
    H_{\mu\nu}H^{\mu\nu}\big] \nn \\
  &&+  \frac{1}{12\pi^2}\int_{\partial M} d^3x\sqrt{-h}[B_1(S)-B_2(S)-
    B_3(S) ],
\end{eqnarray}
where $F=dV$, $H=dS$, $h_{\mu\nu}$ is the induced
metric on the boundary $\partial M$, 
$k_{\mu\nu}=h^{\rho}_{\mu}h^{\sigma}_{\nu}\nabla_{\rho} n_{\sigma}$ is the extrinsic curvature, 
$\bar{k}_{\mu\nu}$ and $k$ denote
the traceless part  and the trace of extrinsic curvatures respectively,
$B_1(S):=\frac{2}{3}k(h^{\mu\nu}+n^{\mu}n^{\nu})S_{\mu}S_{\nu}
+S_n \nabla_{\mu}S^{\mu}+2S_{\mu}h^{\mu\nu}\nabla_n S_{\nu}$,
$B_2(S) :=\frac{1}{3}k S_{\mu}S^{\mu}+n^{\mu}S^{\nu}\nabla_{\nu}S_{\mu}$
and
$B_3(S) := \frac{1}{5} \bar{k}_{\mu\nu}S^{\mu}S^{\nu}$.
Here we choose the normal vector so that $n^{\mu}=-n_{\mu}=(0,-1,0,0)$
in a flat half space.  
The Weyl anomaly (\ref{anomaly}) is Weyl invariant and
satisfies the  Wess-Zumino consistency condition \cite{Wess:1971yu}. 
Note that the bulk contribution to the torsional Weyl anomaly is discussed in \cite{Obukhov:1983mm,Buchbinder:1985ym,Camargo:2022gcw}. To the best of our knowledge, the boundary contribution to the torsional Weyl anomaly (\ref{anomaly}) is new.

We are interested in the expectation value of the chiral current and vector
current in the theory. In 4-dimensions, the
renormalized vacuum expectation value of the chiral current 
derived by the variation of effective action with respect to the
background `axial vector'
\begin{eqnarray} \label{ChiralCurrent}
&&J_S^{\mu}=  \la \bar{\psi} \gamma^{\mu}\gamma_5 \psi \ra
  =\frac{1}{\sqrt{-g}}\frac{\delta I_{\rm eff}}{\delta S_{\mu}}.
\end{eqnarray}
In the following, 
we  show 
that the knowledge of the Weyl
anomaly (\ref{anomaly}) allows one immediately to 
determine \eq{ChiralCurrent} in
closed analytic form.

\section{Chiral Current}

\subsection{Boundary Theory}

Let us first study the chiral current
in 4-dimensional spacetime with
a boundary, say, at $x=0$ of
the coordinate system.
We follow the methods of
\cite{Miao:2017aba,Chu:2018ksb}, where we have studied the expectation
value of current and stress tensor in boundary quantum field theories
\cite{Cardy:2004hm}.
To start with, we note that
since the mass dimension of chiral current is 3,
it takes
the asymptotic form \cite{Deutsch:1978sc}
\begin{eqnarray} \label{Current1}
J_S^{\mu}=\frac{J^{\mu}_0}{x^3}+\frac{J^{\mu}_1}{x^2}+\frac{J^{\mu}_2}{x}+O(\ln x)
\end{eqnarray}
near the boundary. Here $x$ is the proper distance from the boundary, 
$J^{\mu}_{\text{n}}$ have mass dimension $\text{n}$ and depend on only the background
geometry and the background torsion. $J^{\mu}_{\text{n}}$ can be solved
by imposing the conservation law $\nabla_{\mu} J^{\mu}_S=O(1)$
\cite{Chu:2018ksb}, where $O(1)$ denotes the finite part of the chiral anomaly
which is irrelevant to the divergent part of renormalized current. We obtain
\begin{eqnarray} \label{Current2}
J_0^{\mu}=0, \  \ J_1^{\mu}=\lambda \ h^{\mu\nu}S_{\nu},
\end{eqnarray}
where $\lambda$ is some constant. The key point in the above derivations is
that the leading term of chiral current cannot be proportional to the
normal vector $n^{\mu}$, otherwise it cannot satisfy the conservation
law $\nabla_{\mu} (n^{\mu}/x^3)\sim 1/x^4 \ne O(1)$.  We note that unlike the case
of gauge field \cite{Chu:2018ksb},
the transformation $\delta S_{\mu}=\partial_{\mu} \alpha$
changes the torsion and is not required to be a symmetry of the theory,
therefore a non-vanishing $J_1^\mu$ term as in \eq{Current2} is allowed.


Now let us use the ``integrability'' relation \eq{key} for the
renormalized chiral current $J^\m_S$.
To proceed, let us
employ the Gauss normal coordinates to express the metric
$ds^2=-dx^2+ (h_{ij}-2x k_{ij}+ \cdots )dy^i dy^j $ and expand 
$S_{\mu}(x)=S_{0\ \mu}+ x
S_{1\ \mu}+O(x^2)$, where $x\in [0,+\infty)$ and $S_i$
  give the $i^{th}$ derivatives of $S$ at $x=0$.
  Substituting \eq{anomaly},\eq{Current1}, \eq{Current2} into \eq{key},
  after some calculations we obtain
 the chiral current near the boundary:
\begin{eqnarray}
\label{chiralcurrentbdy}
 &&J^{a}_S= \frac{S_0^a}{6 \pi^2 x^2}+\frac{2k^{a}{}_bS^b_0+k S^a_0}{10 \pi^2 x}
+O(\ln x), \nn\\
&&J^{n}_S= \frac{(D_aS^a)_0}{6 \pi^2 x}+O(\ln x), \ \quad x \sim 0,
\end{eqnarray}
where $n$ and $a$ denote respectively the normal and tangential directions, 
$D_a$ is the covariant derivative on the boundary,
and we denote for any function $F(x)$ that  $F_0 = F(x=0)$.
A couple of remarks are in order.
We note that as in the discussion
\cite{Chu:2018ksb,Miao:2017aba}, the total chiral current is finite
since there are boundary contributions to
the chiral current which cancel the divergence from the bulk contribution
(\ref{chiralcurrentbdy}). 
We note that
the result \eq{chiralcurrentbdy} applies not only to conformal
field theory (CFT) but also the general quantum field theory (QFT)
since the Weyl anomaly is well-defined for general quantum field
theories \cite{Duff:1993wm,Brown:1976wc}.
We remark that (\ref{chiralcurrentbdy}) can be verified
by the Green's function method \cite{Hu:2020puq,Chu:2020gwq}.
In a flat half space with $k_{ab}=0$, the correction of Green
function due to torsion is given by
\begin{eqnarray}\label{Gcper}
  G_c(x, x')&=&-\int_Md^4y\sqrt{|g|} G_0(x,y)\gamma^{\mu}\gamma_5
  S_{\mu}(y) G_0(y,x')  +O(S^2)
\end{eqnarray} 
where $G_0$ is Green's function without torsion.  From (\ref{Gcper}), we can
obtain the chiral current by
\begin{eqnarray}\label{regCurrentDirac}
  J^{\mu}_S(x)=-i\lim_{x'\to x} {\rm Tr}_{\rm reg}
  \Big[\gamma^{\mu}\gamma_5 G_c(x,x'))\Big],
\end{eqnarray}
which agrees with (\ref{chiralcurrentbdy}).
Here ${\rm Tr}_{\rm reg}$  means we have  subtracted the reference current
without boundaries. 
Another interesting remark is about the universal nature of the boundary
current \eq{chiralcurrentbdy}. In \cite{Ferreiros:2020uda}
it was shown that torsion does not lead to new chiral 
transport effects
in the bulk since
the response to torsion can be viewed as a manifestation of the chiral vortical 
effect. To see this, it was noted that in the presence of torsion,
the chiral anomaly receives
in addition to the electromagnetic contribution
a torsion contribution
\begin{eqnarray}\label{typeIcurrent-5}
  \partial_{\mu}J_S^{\mu}-T^{\lambda}_{\lambda\mu} J_S^{\mu}
  =c_F \epsilon^{\mu\nu\rho\sigma}F_{\mu\nu}F_{\rho\sigma}+
 c_T \epsilon^{\mu\nu\rho\sigma} \eta_{ab} T^{a}_{\mu\nu}T^{b}_{\rho\sigma}, 
\end{eqnarray}
where the last term denotes the famous Nieh-Yan term (here
we have
focused on the space without curvatures for simplicity)
and the coefficients $c_F$ and $c_T$ have
mass dimensions zero and two respectively.
As a result, unlike $c_F$, $c_T$ depends on regularization and can
be removed by a suitable local counterterm in the background fields
\cite{Chernodub:2021nff}. 
Thus the Nieh-Yan anomaly is not independent, and can be considered a 
manifestation of the chiral anomaly  \cite{Ferreiros:2020uda}.
In fact the form of current non-conservation will depend on the precise 
definition of the current  \cite{Ferreiros:2020uda}. Moreover, a choice of current that is 
based on physical symmetries was suggested and it was shown that the 
Neh-Yan anomaly does not appear  \cite{Ferreiros:2020uda}.

Remarkably, the  boundary chiral current induced by the Weyl anomaly (17) is different as, unlike the effect of torsion on the bulk current in (24), the effect of torsion on the boundary current (17) is non-removable.
Note that all the coefficients of Weyl anomaly
(\ref{anomaly}) are dimensionless. As a result, the Weyl-anomaly
induced chiral current is universal near the boundary, and cannot be
removed by local counterterms.
The universality of the current \eq{chiralcurrentbdy} near the boundary does
not contradict the non-universality of torsion-induced chiral current
in the bulk.
It arises from novel boundary effect that is independent of
renormalization scheme.

\begin{figure}
 \label{dislocation}
\centering
\includegraphics[width=0.4\linewidth]{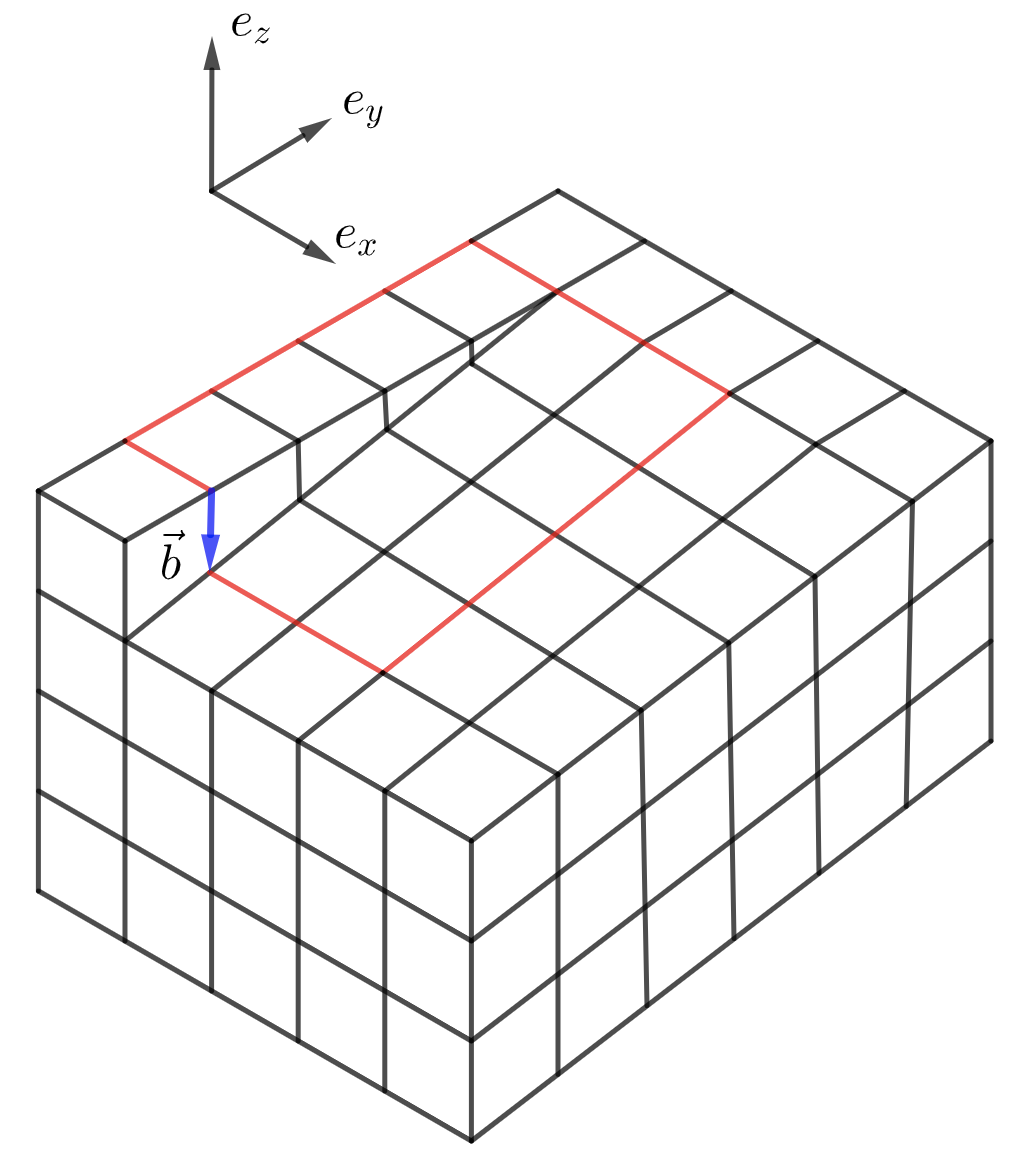}
\caption{Chiral current $\vec{J}_S\sim \vec{b}/x^2$ induced by Screw
  dislocation, where $\vec{b}$ is the Burgers vector.} 
\end{figure}

Let us
briefly comment on how the chiral current (\ref{chiralcurrentbdy}) may be
measured in Dirac
semimetals. As
 shown in Fig. 1, we perform the Screw dislocation of lattices so that
 the red parallelogram does not close, and the missing part is defined
 by the blue Burgers vector.  The density of the Burgers vector
 $\vec{b}$ behaves as the axial vector in Dirac and Weyl semimetals
 \cite{deJuan:2010zz}. 
 From (\ref{chiralcurrentbdy}) together with $\vec{S} \sim \vec{b}$ , 
 we draw conclusions that the Screw dislocation induces an anomalous chiral
 current near the boundary in Dirac and Weyl semimetals
 \begin{eqnarray}\label{chiral current in experiment}
\vec{J}_S\sim \frac{\vec{b}}{x^2}, \quad   x\gtrsim a
\end{eqnarray}
where $a$ denotes the lattice length, 
and we mainly focus on spatial $ \vec{b}$ in this paper. 
Note that our result (\ref{chiralcurrentbdy}) for the continuum is UV finite and is 
independent of regularization. However, as we go from the 
continuum to a lattice, the details of the lattice will enter in 
general, such as a in (\ref{chiral current in experiment}). In principle, other parameters of the 
system such as mass, temperature, hydrodynamic velocity etc may also 
appear in dimensionless combination and correct the overall coefficient 
of the induced current. 
Note that, to obey the bag
boundary condition (\ref{bagBC}), we should place an insulator on the
boundary of the materials so that no current can flow out of the
boundary $x=0$.

Finally we make a remark for 3-dimensions.
Following the same analysis as for the derivation of \eq{Current1},
\eq{Current2}, the renormalized expectation value of the vector current
$J_V^{\mu}=  \la \bar{\psi} \gamma^{\mu} \psi \ra$ takes the form 
\be
J^\m_V = F^{n\m} (\a_1 + \a_2 \ln x),
\ee
where $\a_1, \a_2$ are constant parameters which are sensitive to the boundary
condition. We note that in dimensions $d<4$, the current is not related to the
Weyl anomaly. Hence the parameters $\a_1, \a_2$ are not determined just by
the central charges, but by
further specific details of theory.


\subsection{Conformally Flat Spacetime}

There are also novel chiral current in 4-dimensional 
conformally flat spacetime without boundaries.
To demonstrate this, let us start by deriving the anomalous
transformation rule for the chiral current. 
Consider the theory \eq{action} with metric and
chiral vector field given by
$(g_{\m\n},S_{\m})$. Due to the anomaly, the renormalized effective
action $I_{\rm eff}$ is not invariant under the Weyl transformation.
Generally, we have \cite{footnote1}
\begin{eqnarray}
\frac{\d}{\d\s} I_{\rm eff}(e^{-2\s} g_{\m\n}) =
\mathcal{A} (e^{-2\s} g_{\m\n})
\end{eqnarray} 
for arbitrary finite $\s(x)$.
This can be integrated to give the effective
action \cite{Wess:1971yu,Cappelli:1988vw,Schwimmer:2010za}. Using the fact that
the anomaly (\ref{anomaly}) is Weyl invariant, we obtain immediately the
transformation rule for the effective action:
\bea\label{Ieff-t}
&& I_{\rm eff}(e^{-2\s} g_{\m\n})=  I_{\rm eff}(g_{\m\n}) + \frac{1}{24\pi^2}\int_M
d^4x\sqrt{-g}
H_{\mu\nu}H^{\mu\nu} \sigma, 
\eea
plus a boundary term
$\frac{1}{12\pi^2}\int_{\partial M}\sqrt{-h}[\frac{-1}{5}
  \bar{k}_{\mu\nu}S^{\mu}S^{\nu}+B_1-B_2]\sigma$,
which we drop in spacetime without boundaries. One can
check that the dilaton effective action satisfies Wess-Zumino consistency
$[\delta_{\sigma_1},\delta_{\sigma_2}] I_{\rm eff}=0$.  Using \eq{Ieff-t},
we obtain finally the transformation rule for the chiral
current (\ref{ChiralCurrent}) under Weyl transformation
$g_{\mu\nu} \to g'_{\mu\nu} = e^{-2\sigma} g_{\m\n}$, $S_{\mu}\to S'_{\mu}=S_{\mu}$
\bea \label{chiralcurrentconflat}
J_S^{\mu} = \frac{1}{6\pi^2}
   \nabla_{\nu}(H^{\n\m}\sigma), 
  \eea
  plus a trivial term $e^{-4\sigma} J_S'^{\mu} $.
  Here $J_S^{\mu}$ (resp.  $J'_S{}^{\mu}$)
  denotes the vev of the chiral current of
  the theory \eq{action} in the background spacetime $g_{\mu\n}$
  (resp. $g'_{\mu\n}$).
Taking $g'_{\mu\nu}$ to be the flat spacetime metric and assuming that the
chiral current vanishes in some region of the flat spacetime,
we finally obtain \eq{chiralcurrentconflat} as the
chiral current in conformally flat spacetime 
\be\label{cf-metric}
ds^2 = e^{2 \s} \eta_{\m\n}dx^\m dx^\n.
\ee

Note that the conformal factor $\sigma$
in (\ref{chiralcurrentconflat}) is arbitrary and needs not to be
small. Therefore we can use (\ref{chiralcurrentconflat}) to
calculate the current in general conformally flat spacetimes
such as Anti-de-Sitter space, de-Sitter space and Robertson-Walker
universe. For Robertson-Walker universe
$ds^2=dt^2-a(t)^2(dx^2+dy^2+dz^2)$, 
we have at time $t=t_*$
\begin{eqnarray} \label{chiralcurrentconflat1}
J_S^{\mu}= \frac{1}{6\pi^2} H^{0\mu} H
\end{eqnarray}
where  $H=\dot{a}/a$ is the Hubble parameter.
For simplicity we have chosen $a(t_*)=1$.
In materials, curvature and torsion can be mimicked by disclinations
and dislocations, respectively. Thus, one may measure the effect
(\ref{chiralcurrentconflat}) in Dirac semimetals with suitable
disclinations and dislocations.

\section{Weyl Fermions} 

So far we have focused on Dirac fermions.  The
discussions can be generalized to Weyl fermions straightforwardly.
The real part of Weyl anomaly for Weyl fermions is half of that of
Dirac fermions (\ref{anomaly}). As a result, the anomalous chiral
current is also half of that Dirac fermions
\eq{chiralcurrentbdy}, \eq{chiralcurrentconflat}.  The imaginary
part of Weyl anomaly is parity odd and it is controversial whether
such term exists \cite{Bastianelli:2016nuf,Bonora:2014qla}.  This
imaginary part implies that the theory is non-unitary or there is
absorption and dissipation in materials. For simplicity, let us take
the vector parts of torsion $V_{\mu}$ as an example.  The discussion
for axial vector parts of torsion $S_{\mu}$ is similar.  The Weyl
anomaly of Weyl fermions related to $V_{\mu}$ is
\begin{eqnarray} \label{anomalyWeylfermion}
  \mathcal{A}=\frac{1}{48\pi^2}\int_M d^4x\sqrt{-g}
  \big[F_{\mu\nu}F^{\mu\nu}+ i \frac{3}{2}\eta F_{\mu\nu} {}^*F^{\mu\nu}\big]
  \end{eqnarray}
where ${}^*F^{\mu\nu}=\frac{1}{2}\epsilon^{\mu\nu\alpha\beta}F^{\alpha\beta}$,
$\eta=0\ {\rm or}\ 1$ denote the controversy. Following the above approach,
we derive the currents
\begin{eqnarray}\label{currentWeylfermionbdy}
  J_V^{\mu} =
  \frac{ n_{\nu}F^{\nu\mu}+i \frac{3}{2}\eta\
    n_{\nu}{}^*F^{\nu\mu} }{12\pi^2 x} + O(\ln x), \end{eqnarray}
near a boundary and 
\begin{eqnarray}\label{currentWeylfermionconflat}
J_V^{\mu} = e^{-4\sigma}J_V'^{\mu}+\frac{1}{12\pi^2} \nabla_{\nu}\Big(
  F^{\n\m}\sigma+i \frac{3}{2}\eta \ {}^*F^{\n\m}\sigma\Big), 
\end{eqnarray}
in a conformally flat space without boundaries. 
 Recall that the edge dislocations and screw dislocations can induce effective vectors and axial vectors coupled with fermions in materials, respectively. 
 Thus, the vector $V_{\mu}$ can be realized by either an electromagnetic
 field or suitable edge dislocations in Weyl semimetals. 
  It is interesting
 to measure the predicted current
 \eq{currentWeylfermionbdy}, \eq{currentWeylfermionconflat} in Weyl
 semimetals, which can help to clarify the theoretical controversy
 that if an imaginary Pontryagin density could appear in
 the Weyl anomaly
 \cite{Bastianelli:2016nuf,Bonora:2014qla}.


 \vskip 1cm
 \centerline{******}
 
 Summarizing, we have shown in this paper that, due to Weyl anomaly,
 torsion can lead to novel currents and chiral currents for Dirac and
 Weyl fermions. We propose to measure these interesting effects in
 Dirac and Weyl semimetals with suitable dislocations. These
 experiments can help to clarify the theoretical controversy that if
 an imaginary Pontryagin density could appear in the Weyl anomaly.

\section*{Acknowledgments}

C.S.Chu acknowledge support of this work by NCTS and the grant 110-2112-M-007-015-MY3 of the National Science and Technology Council of Taiwan. 
R.X.Miao acknowledges the supports from Guangdong Basic and Applied Basic Research Foundation (No.2020A1515010900) and National Natural Science Foundation of China (No. 11905297).


\begin{thebibliography}{99}
\bibitem{footnote1}
  We have find it conveniently to consider the Weyl transformation
  $g_{\mu\nu} \to e^{-2\s} g_{\mu\nu}$ here.
\end{thebibliography}
\end{document}